\documentclass[preprint,tighten]{aastex}

\submitjournal{ApJ}
\accepted{5/27/21}

\usepackage[modulo]{lineno}
\usepackage{float}
\usepackage{amsmath}
\usepackage{rotating}
\usepackage{amssymb}
\usepackage{hyperref}
\usepackage[utf8]{inputenc}

\usepackage{color, soul} 

\def\mushroom{PSR~B0355+54}
\def\fgl{3FGL~J0359.5+5413}
\def\fglpsr{PSR~J0359+5414}

\def\morla{PSR~J0357+3205}

\def\psrj{PSR~J1740+1000}

\def\etal{et~al.}

\newcommand\chandra{{\it Chandra}}

\newcommand\xmm{{\it XMM-Newton}}
\newcommand\XMM{{\it XMM-Newton}}
\newcommand\veritas{{\rm VERITAS}}

\newcommand\fermi{{\it Fermi}}

\shorttitle{VERITAS Search for TeV emission from Pulsar Tails}
\shortauthors{VERITAS et al.}

\begin{document}

\title{A Search for TeV Gamma-ray Emission from Pulsar Tails by VERITAS}




\author{W.~Benbow}\affiliation{Center for Astrophysics $|$ Harvard \& Smithsonian, Cambridge, MA 02138, USA}
\author{A.~Brill}\affiliation{Physics Department, Columbia University, New York, NY 10027, USA}
\author{J.~H.~Buckley}\affiliation{Department of Physics, Washington University, St. Louis, MO 63130, USA}
\author{M.~Capasso}\affiliation{Department of Physics and Astronomy, Barnard College, Columbia University, NY 10027, USA}
\author{A.~J.~Chromey}\affiliation{Department of Physics and Astronomy, Iowa State University, Ames, IA 50011, USA}
\author{M.~Errando}\affiliation{Department of Physics, Washington University, St. Louis, MO 63130, USA}
\author{A.~Falcone}\affiliation{Department of Astronomy and Astrophysics, 525 Davey Lab, Pennsylvania State University, University Park, PA 16802, USA}
\author{K.~A~Farrell}\affiliation{School of Physics, University College Dublin, Belfield, Dublin 4, Ireland}
\author{Q.~Feng}\affiliation{Department of Physics and Astronomy, Barnard College, Columbia University, NY 10027, USA}
\author{J.~P.~Finley}\affiliation{Department of Physics and Astronomy, Purdue University, West Lafayette, IN 47907, USA}
\author{G.~M.~Foote}\affiliation{Department of Physics and Astronomy and the Bartol Research Institute, University of Delaware, Newark, DE 19716, USA}
\author{L.~Fortson}\affiliation{School of Physics and Astronomy, University of Minnesota, Minneapolis, MN 55455, USA}
\author{A.~Furniss}\affiliation{Department of Physics, California State University - East Bay, Hayward, CA 94542, USA}
\author{A.~Gent}\affiliation{School of Physics and Center for Relativistic Astrophysics, Georgia Institute of Technology, 837 State Street NW, Atlanta, GA 30332-0430}
\author{C.~Giuri}\affiliation{DESY, Platanenallee 6, 15738 Zeuthen, Germany}
\author{D.~Hanna}\affiliation{Physics Department, McGill University, Montreal, QC H3A 2T8, Canada}
\author{T.~Hassan}\affiliation{DESY, Platanenallee 6, 15738 Zeuthen, Germany}
\author{O.~Hervet}\affiliation{Santa Cruz Institute for Particle Physics and Department of Physics, University of California, Santa Cruz, CA 95064, USA}
\author{J.~Holder}\affiliation{Department of Physics and Astronomy and the Bartol Research Institute, University of Delaware, Newark, DE 19716, USA}
\author{G.~Hughes}\affiliation{Center for Astrophysics $|$ Harvard \& Smithsonian, Cambridge, MA 02138, USA}
\author{T.~B.~Humensky}\affiliation{Physics Department, Columbia University, New York, NY 10027, USA}
\author{W.~Jin}\affiliation{Department of Physics and Astronomy, University of Alabama, Tuscaloosa, AL 35487, USA}
\author{P.~Kaaret}\affiliation{Department of Physics and Astronomy, University of Iowa, Van Allen Hall, Iowa City, IA 52242, USA}
\author{Oleg Kargaltsev}\affil{Department of Physics, The George Washington University, Washington, DC 20052, USA}
\author{M.~Kertzman}\affiliation{Department of Physics and Astronomy, DePauw University, Greencastle, IN 46135-0037, USA}
\author{D.~Kieda}\affiliation{Department of Physics and Astronomy, University of Utah, Salt Lake City, UT 84112, USA}
\author{Noel Klingler}\affil{Department of Astronomy and Astrophysics, The Pennsylvania State University, University Park, PA 16802, USA}
\author{S.~Kumar}\affiliation{Physics Department, McGill University, Montreal, QC H3A 2T8, Canada}
\author{M.~J.~Lang}\affiliation{School of Physics, National University of Ireland Galway, University Road, Galway, Ireland}
\author{M.~Lundy}\affiliation{Physics Department, McGill University, Montreal, QC H3A 2T8, Canada}
\author{G.~Maier}\affiliation{DESY, Platanenallee 6, 15738 Zeuthen, Germany}
\author{C.~E~McGrath}\affiliation{School of Physics, University College Dublin, Belfield, Dublin 4, Ireland}
\author{P.~Moriarty}\affiliation{School of Physics, National University of Ireland Galway, University Road, Galway, Ireland}
\author{R.~Mukherjee}\affiliation{Department of Physics and Astronomy, Barnard College, Columbia University, NY 10027, USA}
\author{D.~Nieto}\affiliation{Institute of Particle and Cosmos Physics, Universidad Complutense de Madrid, 28040 Madrid, Spain}
\author{M.~Nievas-Rosillo}\affiliation{DESY, Platanenallee 6, 15738 Zeuthen, Germany}
\author{S.~O'Brien}\affiliation{Physics Department, McGill University, Montreal, QC H3A 2T8, Canada}
\author{R.~A.~Ong}\affiliation{Department of Physics and Astronomy, University of California, Los Angeles, CA 90095, USA}
\author{A.~N.~Otte}\affiliation{School of Physics and Center for Relativistic Astrophysics, Georgia Institute of Technology, 837 State Street NW, Atlanta, GA 30332-0430}
\author{S.~Patel}\affiliation{Department of Physics and Astronomy, University of Iowa, Van Allen Hall, Iowa City, IA 52242, USA}
\author{K.~Pfrang}\affiliation{DESY, Platanenallee 6, 15738 Zeuthen, Germany}
\author{M.~Pohl}\affiliation{Institute of Physics and Astronomy, University of Potsdam, 14476 Potsdam-Golm, Germany and DESY, Platanenallee 6, 15738 Zeuthen, Germany}
\author{R.~R.~Prado}\affiliation{DESY, Platanenallee 6, 15738 Zeuthen, Germany}
\author{J.~Quinn}\affiliation{School of Physics, University College Dublin, Belfield, Dublin 4, Ireland}
\author{K.~Ragan}\affiliation{Physics Department, McGill University, Montreal, QC H3A 2T8, Canada}
\author{P.~T.~Reynolds}\affiliation{Department of Physical Sciences, Munster Technological University, Bishopstown, Cork, T12 P928, Ireland}
\author{D.~Ribeiro}\affiliation{Physics Department, Columbia University, New York, NY 10027, USA}
\author{G.~T.~Richards}\affiliation{Department of Physics and Astronomy and the Bartol Research Institute, University of Delaware, Newark, DE 19716, USA}
\author{E.~Roache}\affiliation{Center for Astrophysics $|$ Harvard \& Smithsonian, Cambridge, MA 02138, USA}
\author{J.~L.~Ryan}\affiliation{Department of Physics and Astronomy, University of California, Los Angeles, CA 90095, USA}
\author{M.~Santander}\affiliation{Department of Physics and Astronomy, University of Alabama, Tuscaloosa, AL 35487, USA}
\author{G.~H.~Sembroski}\affiliation{Department of Physics and Astronomy, Purdue University, West Lafayette, IN 47907, USA}
\author{R.~Shang}\affiliation{Department of Physics and Astronomy, University of California, Los Angeles, CA 90095, USA}
\author{Igor Volkov}\affil{Department of Physics, The George Washington University, Washington, DC 20052, USA}
\author{S.~P.~Wakely}\affiliation{Enrico Fermi Institute, University of Chicago, Chicago, IL 60637, USA}
\author{A.~Weinstein}\affiliation{Department of Physics and Astronomy, Iowa State University, Ames, IA 50011, USA}
\author{P.~Wilcox}\affiliation{School of Physics and Astronomy, University of Minnesota, Minneapolis, MN 55455, USA}
\author{D.~A.~Williams}\affiliation{Santa Cruz Institute for Particle Physics and Department of Physics, University of California, Santa Cruz, CA 95064, USA}

\begin{abstract}

We report on the search for very-high-energy gamma-ray emission from the regions around three nearby supersonic pulsars (\mushroom, \morla\ and \psrj) that exhibit long X-ray tails. To date there is no clear detection of TeV emission from any pulsar tail that is prominent in X-ray or radio.   We provide upper limits on the TeV flux, and luminosity, and also compare these limits with other pulsar wind nebulae detected in X-rays and the tail emission model predictions. We find that at least one of the three tails is likely to be detected in observations that are a factor of 2-3 
 more sensitive. The analysis presented here also has implications for deriving the properties of pulsar tails, for those pulsars whose tails could be detected in TeV.  

\end{abstract}


\section{Introduction}\label{sec:intro}

Pulsars are among nature's most powerful particle accelerators, capable of producing particles with energies up to a few PeV.  
As a neutron star rotates, it imparts some of its immense rotational energy into a magnetized ultra-relativistic particle wind. 
Although the wind bulk flow speed is initially highly relativistic, the flow decelerates abruptly at the termination shock (TS) to only mildly-relativistic speed, due to interaction with the surrounding medium.
Beyond the TS, the pitch angle distribution of the wind particles  becomes more isotropic and their  synchrotron radiation becomes observable, from radio to gamma-rays, as a pulsar wind nebula (PWN, plural: PWNe). 

In addition to pulsar properties (spin-down energy-loss rate $\dot{E}$, viewing angle $\zeta$, magnetic-spin axis offset $\alpha$, etc.), PWN sizes and morphologies depend upon the properties of the interstellar medium (ISM), such as the ambient pressure and local speed of sound, $c_s$. 
Since the ``birth kicks'' that pulsars receive in their progenitor supernova explosions, $v_p$, are typically on the order of a few hundred km s$^{-1}$ \citep{Verbunt2017}, pulsars will only remain in their host supernova remnant (SNR) for a few tens of kyr. 
Once a pulsar leaves its SNR, it traverses an environment with a much lower speed of sound: in the ISM, $c_s$ typically ranges from a few to a few tens of km s$^{-1}$.
Consequently, most pulsars outside of their SNRs move supersonically (i.e., $v_p / c_s \equiv \mathcal{M} \gg 1$, where $\mathcal{M}$ is the Mach number).
For supersonic pulsars, the ram pressure exerted upon the shocked pulsar wind (PW) by the oncoming ISM confines the shocked PW to the direction opposite that of the pulsar motion, forming a bow shock PWN ``head'' and an extended pulsar ``tail''. 

Thus, pulsar tails are ram-pressure-confined outflows of the magnetized synchrotron-emitting relativistic particle wind.
Radio and X-ray observations of pulsar tails have shown that they can extend distances of $\sim10$ pc behind their pulsars (see, e.g., \citealt{Kargaltsev2017} and \citealt{Bykov2017} for reviews), with the longest being the Mouse and Frying Pan PWNe, which both span about 18 pc in radio (see Figure 3 of \citealt{Klingler2018}, and \citealt{Ng2012}).  We also note that, in some cases, the confinement of the particle wind by the ram pressure may be associated with the relative motion of the surrounding medium, rather than with the pulsar's motion. For example,  the analysis of radio observations \citep{Chevalier2011} suggests that for the Vela PWN the  flow of the surrounding medium is mildly supersonic, after recent passage of the reverse shock and could have been even more supersonic during the passage. H.E.S.S.\ observations \citep{Aharonian2006,Abramowski2012} revealed a strongly elongated TeV structure south of the Vela pulsar (which is also bright in radio) whose X-ray emission has both thermal and non-thermal components \citep{Slane2018}, of which the latter may belong to the PWN  (as discussed in \citealt{Bykov2017}).


PWNe typically emit most of their energy via synchrotron radiation from radio to MeV $\gamma$-rays, but PWNe can also be prominent sources of GeV and TeV $\gamma$-rays, through the upscattering of CMB photons by the ultra-relativistic wind particles. 
The higher energy particles which emit synchrotron radiation in X-rays cool on relatively short timescales ($\sim 1$ kyr, for typical magnetic fields $B_{\rm PWN} \gtrsim 10$ ${\rm \mu G}$). The lower energy particles which emit inverse-Compton (IC) radiation at GeV/TeV energies may cool on  longer timescales ($\sim 10 - 100$ kyr, for typical ISM photon densities) if the magnetic field  strength drops to few $\mu$Gauss farther away from the pulsar.  Therefore,  PWNe can  extend to  larger scales in the TeV regime.
In fact, most identified Galactic TeV sources are believed to be PWNe (and/or the SNRs in which they reside; \citealt{Kargaltsev2013}, \citealt{Hui2015}, \citealt{HESS2017}). 
However, puzzlingly, no supersonic PWNe (SPWNe) with long tails seen in radio and X-rays have been detected in TeV energies yet. 
 Although the very nearby ($d\approx250$ pc) Geminga pulsar is supersonic \citep{Posselt2017} and its PWN has been detected in $\gamma$-rays (see \citealt{Abeysekara2017} \& \citealt{Linden2017}), the extended tail has not been detected in X-rays.

Detecting IC TeV  emission from  long tails, whose synchrotron emission has been detected in X-rays, would enable an informative diagnostic of pulsar wind properties, since modeling for these SPWNe is simpler than for PWNe still residing within their host SNRs, where one can expect additional contributions to the TeV emission arising from the interaction between the pulsar wind and reverse shock of the SNR or the pulsar wind and dense material inside the SNR.



\textbf{\morla}\ (hereafter J0357) is a nearby radio-quiet pulsar discovered through a blind frequency search of the \fermi\,-LAT data \citep{Abdo2009}. Its spin-down power ($\dot{E} = 5.9\times10^{33}$ erg s$^{-1}$) is among the lowest of gamma-ray pulsars, even when selecting for pulsars of similar age ($\tau = 540$ kyr). The distance to J0357 remains uncertain, although \citet{DeLuca2011} used a $\gamma$-ray ``pseudo-distance" relation \citep[see e.g.,][]{SazParkinson2010} to estimate a distance of 500~pc (we adopt this distance below).

\begin{figure}
\epsscale{1.15}
\plotone{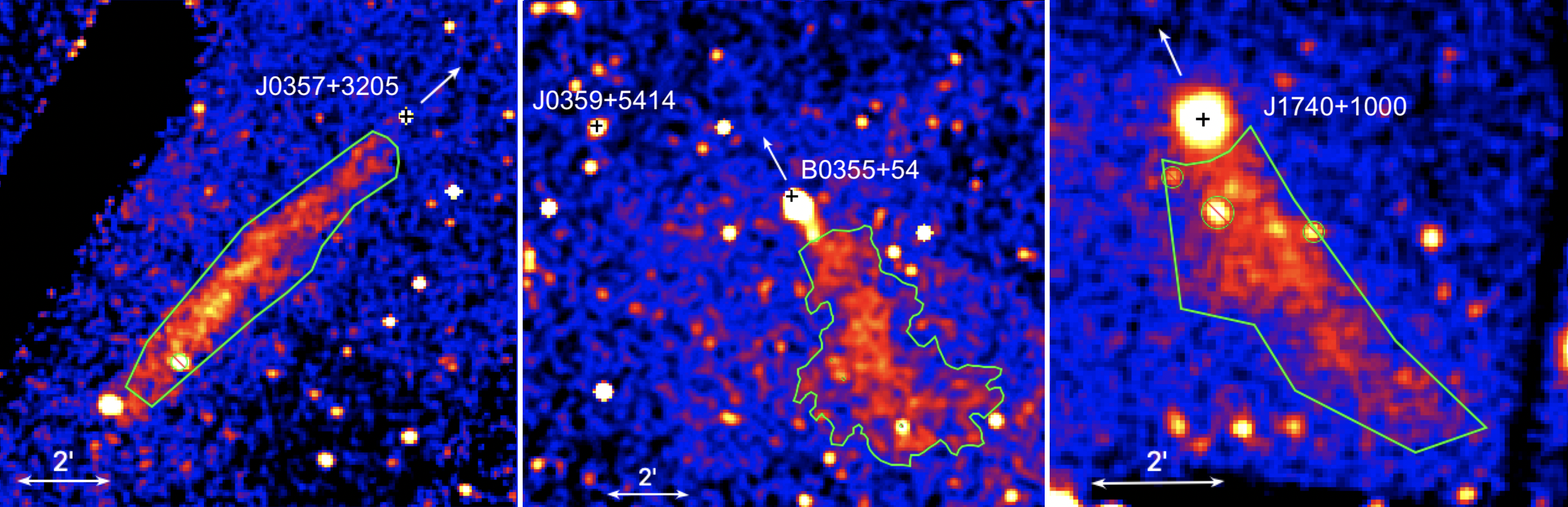}
\caption{The tails of PSRs J0357+3205 (left, \chandra\.), B0355+54 and J0359+5414 (center, \chandra\.), and J1740+1000 (right, {\sl XMM-Newton}).  The X-Ray spectral extraction regions are shown in green, the black crosses mark the pulsar positions, and the white arrows mark their approximate/inferred directions of motion. 
}
\label{fig-pulsar-tails}
\end{figure}

Deep {\sl Chandra X-ray Observatory} ({\sl CXO}) observations revealed an elongated X-ray tail \citep{DeLuca2011} whose emission only becomes visible 20$''$ behind PSR J0357, and extends for an angular distance of $~9'$, reaching maximum brightness $\sim4-7'$ away from the pulsar (see Figure \ref{fig-pulsar-tails}). 
No significant energy-dependent changes in morphology were seen.
The strange morphology of the diffuse emission (i.e., the lack of a bright compact PWN head near the pulsar) is at odds with the morphologies of most other PWNe resolved by the {\sl CXO}.  
One possible explanation could be that the compact PWN is simply unresolved due to its small angular size. 
This would imply a very large pulsar velocity.

Follow-up observations with \chandra\ revealed a proper motion of the pulsar well aligned with the main axis of the nebula, with a transverse velocity of $\sim390 (d/500$ pc) km~s$^{-1}$ \citep{DeLuca2013}. Although the 3D velocity could be much higher, a large angle between the velocity vector and the sky plane would imply a significantly longer tail extent than that seen in projection.  
Similar to J1740, below, the pulsar is located far from the Galactic plane ($b = -16\degr$) with a proper motion angled slightly towards the plane (inclination $\sim2\degr$). 

\textbf{\mushroom}\ (J0358+5413; hereafter B0355) is a radio-loud pulsar with characteristic age $\tau =  564$ kyr and spin-down power $\dot{E} = 4.5\times10^{34}$ erg s$^{-1}$. Using very long baseline interferometry, \citet{Chatterjee2004} derived a parallax-determined distance of $1.04^{+0.21}_{-0.16}$~kpc and transverse proper velocity of $61^{+12}_{-9}$~km~s$^{-1}$. \citet{Verbiest2012} applied a Lutz-Kelker correction to this measurement, giving a distance of $1.0^{+0.2}_{-0.1}$~kpc.
 
\citet{McGowan2006} reported on the analysis of the 66 ks {\sl CXO} ACIS observation which resolved bright compact emission extending up to $50''$ from the pulsar, dubbed the `compact nebula' (CN), and a fainter $7'$-long tail, both extending in the direction opposite the pulsar's direction of motion. 
Using a deeper (395 ks) set of subsequent \chandra\ observations, \citet{Klingler2016b} reported a CN photon index of $\Gamma=1.54\pm0.05$, and found that the tail exhibits 
no spectral cooling across its visible extent, with a best-fit $\Gamma=1.74\pm0.08$ for the entire tail (excluding the CN).

The only source from the 3rd \fermi-LAT catalog (3FGL; \citealt{Acero2015}) within 2\degr\ of B0355 is \fgl, offset from the pulsar's position by $5.5'$. 
\fgl\ was identified as a pulsar, J0359+5414 (hereafter J0359), in the Einstein@Home Gamma-Ray Pulsar Survey \citep{Clark2017}.
J0359 is a relatively young ($\tau=P/2\dot{P}=75.2$ kyr), energetic ($\dot{E}=1.3\times10^{36}$ erg s$^{-1}$), radio-quiet pulsar with period $P=79$ ms, which exhibits a single broad peak in its $\gamma$-ray light curve (see \citealt{Clark2017}).
The deep 395 ks of ACIS-I observations (ObsIDs 14688--14690, 15548--15550, 15585, and 15586) of B0355 contain J0359 in their field of view.


\textbf{\psrj}\ (hereafter J1740) is a young pulsar (characteristic age $\tau = 114$~kyr, spin-down power $\dot{E} = 2.3\times10^{35}$ erg s$^{-1}$) discovered well above the Galactic plane ($b = 20.4\degr$) by the Arecibo Telescope~\citep{McLaughlin2000}. \citet{McLaughlin2002} saw no filamentary or bow-shock structures with the Very Large Array at 1.4 GHz, and noted that a high spatial velocity (oriented perpendicular to the Galactic plane) of $v_z > 4100\ \mathrm{km\ s^{-1}}$ would be required for the pulsar to reach its current Galactic latitude within its characteristic age, if it was born on the Galactic plane at the dispersion measure distance of $\sim1.4$ kpc.\footnote{However, a velocity of $\sim$4000 km s$^{-1}$ is unrealistically high.  The average 3D pulsar velocity has been found to be $400\pm40$ km s$^{-1}$ \citep{Hobbs2005}, with the highest velocities being 765 km s$^{-1}$ (PSR B2224+65; \citealt{Deller2019}) and $\sim$1000--2000 km s$^{-1}$ (inferred; PSR J1101--6101; \citealt{Pavan2014}). 
This implies that this pulsar was born above the Galactic plane, or that its true age is a factor of a few older than its spin-down age.} Using a pair of {\sl CXO} observations with a 10-year baseline, \citet{Halpern2013} set an upper limit on the transverse proper motion which implies a displacement of $<2\degr$ within the characteristic age, supporting the premise that the progenitor star either originated in the halo or escaped from the disk long before the pulsar was born.

\citet{Kargaltsev2008} reported the discovery of an extended (5.5$'$, possibly up to 7$'$) X-ray tail behind J1740 with an \xmm\ observation (46.4~ks effective exposure). A power-law (PL) spectral fit to the tail emission showed a relatively hard photon index ($\Gamma=1.4\pm0.2$), but high background rates and the limited angular resolution of \xmm\ resulted in large uncertainties. 
The orientation of the tail implies that the pulsar is moving at an angle of $\sim7\deg$ towards the Galactic plane, further corroborating a halo-star progenitor.
No $\gamma$-ray emission (pulsed nor unpulsed) from J1740 is seen with the {\sl Fermi}-LAT.


\bigskip

In terms of their spin-down properties, J1740 and J0359 are typical representatives of the pulsar population associated with TeV sources (see the reviews by \citealt{Kargaltsev2010,Kargaltsev2013}). The other two pulsars, B0355 and J0357, are older but relatively nearby (see Table ~\ref{tab:psr_parameters} for a summary of the pulsar parameters), so it is reasonable to expect that TeV emission from their relic PWNe could be detected. This prompted us to carry out observations of these fields with the Very Energetic Radiation Imaging Telescope Array System (VERITAS).  

In this paper we present the results of a search for TeV emission from three supersonic pulsars (B0355+54, J0357+3205, and J1740+1000) with VERITAS and compare the results to the predictions of a multiwavelength emission model adopted to the case of SPWNe. 
In Section 2 we describe VERITAS observations and data analysis as well as X-ray data. The results of TeV and X-ray observations are presented in Section 3. In Section 4 we discuss the implications of the results in the context of a simple multiwavelength emission model of a SPWN (the model details are given in the Appendix). We conclude by summarizing our findings in Section 5.  

\begin{table}[htbp]
\begin{center}
\begin{tabular}{lcccc}
\hline\hline
NAME & J0357+3205 & B0355+54 & J0359+5414 & J1740+1000 \\
\hline
R.A.\ & 03:57:52.5 & 03:58:53.7 & 03:59:26.0 & 17:40:25.9 \\
Decl.\ & +32:05:25 & +54:13:13.7 & +54:14:55.7 & +10:00:06.3 \\
$\mu_\alpha$ (mas yr$^{-1}$) & 117 & 9.3 & -- & -- \\
$\mu_\delta$ (mas yr$^{-1}$) & 115 & 8.3 & -- & -- \\
Gal.\ longitude (deg) & 162.76 & 148.19 & 148.231 & 34.011 \\
Gal. latitude (deg) & -16.006 & 0.811 & 0.883 & 20.268 \\
Spin period, $P$ (ms) & 444.1 & 156.4 & 79.4 & 154.1 \\
Period derivative ($10^{-14}$), $\dot{P}$ & 1.30 & 0.439 & 1.67 & 2.15 \\
Dispersion measure (pc cm$^{-3}$) & -- & 57.14 & -- & 23.9 \\
Distance (kpc) & 0.83 & 1.04 & -- & 1.23 \\
Spin-down age, $\tau_{\rm sd} = P/(2\dot{P})$ (kyr) & 540 & 564 & 75.2 & 114 \\
Spin-down power, $\dot{E}$ (erg s$^{-1}$) & $5.88\times10^{33}$ & $4.54\times10^{34}$ & $1.32\times10^{36}$ & $2.32\times10^{35}$ \\
Surface magnetic field, $B_s$ ($10^{12}$) & 2.43 & 0.839 & 1.2 & 1.84 \\
\hline\hline
\end{tabular}
\end{center}
\caption{Pulsar parameters (from the ATNF Pulsar Catalog, \citealt{Manchester2005}).}
\label{tab:psr_parameters} 
\end{table}

\begin{figure}
\epsscale{1.0}
\plotone{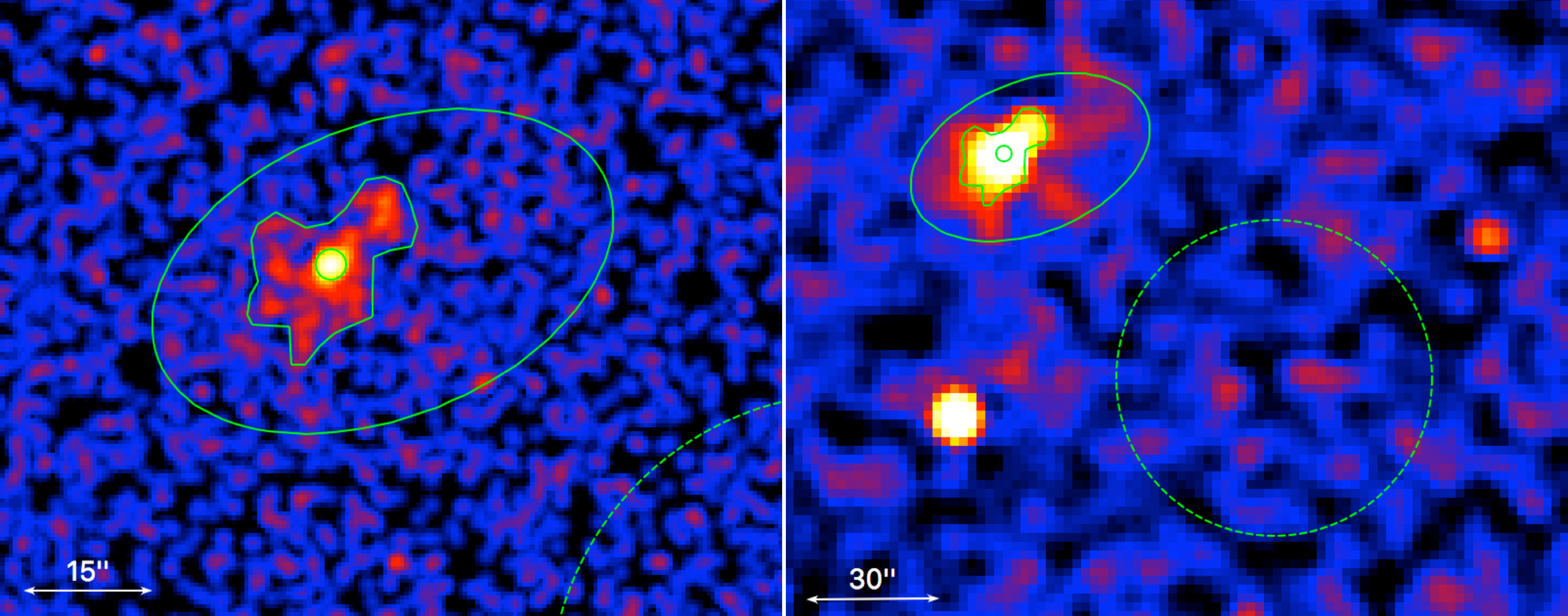}
\caption{Deep \chandra\ ACIS-I (0.5--8 keV) images of J0359.  {\sl Left:} Image smoothed with an $r=1\farcs48$ Gaussian kernel.  {\sl Right:} image binned by a factor of 4, and smoothed with an $r=5\farcs9$ Gaussian kernel to better show the extended emission.  The region 
shown by the $r=1\farcs5$ circle is used to exclude the pulsar. The compact nebula (CN) region is shown by the polygon (excluding the pulsar region), and the extended nebula emission region is shown by the ellipse (excluding the CN region).  The dashed circle shows the region used for background subtraction.}
\label{fig-J0359}
\end{figure}


\section{Data Analysis}\label{sec:observations}

\subsection{VERITAS Instrument and Observations}

\veritas\ is comprised of four imaging atmospheric Cherenkov telescopes (IACTs) operating at the Fred Lawrence Whipple Observatory (FLWO) in southern Arizona, USA. Cherenkov light flashes from gamma-ray and cosmic-ray induced air showers are reflected from 12-meter Davies-Cotton reflectors onto the telescope cameras. Each camera has a field-of-view of 3.5\degr\ and is composed of 499 photomultiplier tubes (PMTs). A three-level trigger requires coincident signals from at least two telescopes to initiate the array-wide read-out of the PMT signals \citep{Holder2006}. The instrument is sensitive over the energy range 0.085--30 TeV, with a typical energy resolution of 20\% and angular resolution of 0.1\degr\ at 1~TeV \citep{Park2015}.

Observations of each source were taken in `wobble' mode, where the telescope pointing is offset from the nominal source position to allow for simultaneous exposure of the source and background regions. The exposures were taken using an angular offset of 0.5\degr\ around the pulsar position, with the offset direction alternating between the four cardinal directions for each 20-30 minute run. \veritas\ observed J1740 in 2008, April to May, resulting in 12.8 hours of livetime after performing quality selection to remove periods of poor weather and/or hardware performance. Observations of J0357 were taken in 2011 September to December, with selected livetime of 8.5 hours. Initial observations of B0355 were taken in 2014 September to October, with selected livetime of 10.2 hours. An additional 12.1 hours were taken in 2015 September to November, with the pointings wobbling around a position in the middle of the X-ray tail (RA/Dec: 59.6864, 54.1755).

The data were processed using standard \veritas\ analysis techniques. Independent analyses were conducted using two software packages, as described by \citet{Daniel2008} and \cite{2017ICRC...35..747M}. The shower images are parameterized by their principal moments \citep{Hillas1985}, with the shower direction and impact parameter calculated using the stereoscopic views from multiple telescopes \citep{Hofmann1999}. Parameters for separating gamma-ray and cosmic-ray induced showers, as well as the energy of the initial particle, are estimated by comparison with Monte-Carlo simulation \citep[see e.g.][]{Aharonian1997, Krawczynski2006}.

Three standard sets of gamma-ray selection criteria (`cuts') were used, optimized for sources with soft, moderate, or hard spectra (see \cite{Park2015} for details on VERITAS performance over the different epochs of the array). 
In the search for a possible extension in the very-high-energy (VHE; $E>100$~GeV) emission, the following  values were chosen for the radius of the source integration region ($\theta$): 0.1\degr, optimized for point-source searches (0.173\degr\ for soft cuts), and 0.235\degr\ for extended emission.  
An \textit{a priori} search region was also defined for each source, as the VHE emission could originate from a particle population which is older, and thus further downstream, than that which gives rise to the X-ray tail.  To account for possible broadening of each tail at large offsets from the pulsar, a symmetric trapezoidal region was used, aligned with the X-ray tail. The areas were 0.4\degr\  wide at a distance 0.2\degr\ ahead of the pulsar and extended back along the tail with an opening angle of 20\degr, terminating at a distance of 0.2\degr\ beyond the TeV tail, which was assumed to be 4 times as long as the X-ray tail (c.f. the quoted VERITAS angular resolution of 0.1\degr\ ). 
In the search for VHE emission in the region of interest, the background was estimated using the ring-background method \citep{Berge2007} and the statistical significance of the excess was calculated using Equation~17 of \citet{Li1983}.

\subsection{\chandra\ observations of PSR J0357+3205}
We analyzed 134 ks of {\sl Chandra} ACIS-S observations of PSR J0357 (ObsIDs 11239, 12008, 14207, and 14208).
The \chandra\ data were processed using the standard {\tt chandra\_repro} routine from the \chandra\ Interactive Analysis of Observations (CIAO) software package version 4.11, and Calibration Data Base (CALDB) version 4.8.3.  

All spectra were extracted from each observation individually using {\tt specextract} and then combined with {\tt combine\_spectra} (due to the low number of counts in each individual observation).  

\subsection{\chandra\ observations of B0355+54 and  PSR J0359+5414}
We also analyzed the 395 ks of ACIS-I data carried out to study B0355 (ObsIDs 14688--14690, 15548--15550, 15585, 15586), which, serendipitously, also contained PSR J0359+5414 in the ACIS field of view. 
The data were processed and the spectra were extracted and fitted the same way as for PSR J0357+3205.

\subsection{\XMM\ observations of PSR J1740+1000}
We analyzed 532 ks of new {\sl XMM-Newton} data carried out to study PSR J1740 (ObsIDs 0803080201, 0803080301, 0803080401, and 0803080501).\footnote{It is worth noting that a 5th exposure of similar length was also carried out (ObsID 0803080101), however we excluded this observation since the majority of the exposure was heavily contaminated due to periods of high solar flaring and/or particle background.}
We only analyzed data from the EPIC MOS1/2 detectors, since the EPIC pn detector was operated in Small Window mode, and thus, half of the tail lies outside of the detector's small field of view.
The data were processed using the standard {\tt emproc} routine from the {\sl XMM} Science Analysis Software (SAS; version 16.1). 
We manually filtered the data for periods of high background and flaring. 

Spectra from each observation and corresponding response files were extracted and produced using the standard routines,\footnote{see https://www.cosmos.esa.int/web/xmm-newton/sas-thread-mos-spectrum} {\tt evselect}, {\tt arfgen}, and {\tt rmfgen}, and grouped to 125 counts per bin using {\tt specgroup}. 

\bigskip
All X-ray spectra were fitted using XSPEC (version 12.10.1).
In all spectral fits we used a simple absorbed power-law (PL) model (XSPEC's {\tt tbabs(pow)}) which uses absorption cross sections from \citet{Wilms2000}.  
All uncertainties listed correspond to 1$\sigma$.


\section{RESULTS}

\subsection{Upper Limits on VHE Emission}\label{sec:results}


No significant emission was detected within the search regions using any of the analysis cuts. Upper limits (ULs) on the integral gamma-ray fluxes above the threshold energies\footnote{The `energy threshold' is defined as the energy at which the differential rate of reconstructed gamma-rays is at its maximum. It is a good indicator of the energy-scale of the analysis, but this threshold is higher than the trigger energy threshold, meaning that gamma-rays below the `energy threshold' are present in the final data.} were calculated using the method of \citet{Rolke2005} at a 95\% confidence level (CL). A bounded confidence interval was used to avoid the unphysical case of negative source counts. A power-law source spectrum was assumed, with an index $\Gamma = 2.5$. With no other \textit{a priori} test positions in the search regions, the limits were taken at positions centered on the middle of the X-ray tails. The backgrounds were estimated using the reflected-regions method \citep{Berge2007}.
Tables~\ref{tab:veritas_upperlimits_morla},  \ref{tab:veritas_upperlimits_mushroom}, and \ref{tab:veritas_upperlimits_j1740}, list the results for each source.

\begin{table}[htbp]
\begin{center}
\begin{tabular}{lcccccc}
\hline\hline
Spectral cuts	& $\theta$		& $\mathrm{E_{threshold}}$	& Significance & $95\%$ CL flux UL				& $95\%$ CL flux UL	& $95\%$ CL luminosity UL  \\
		&			&				& (Pre-Trials)	&($>\mathrm{E_{threshold}}$)			&(1--10~TeV)		&(1--10~TeV) \\
		& [\degr]		&  [GeV]			& [Sigma]	& [$\mathrm{10^{-13}\ cm^{-2}\ s^{-1}}$] 	& [$\mathrm{10^{-13}\ cm^{-2}\ s^{-1}}$]  & [$\mathrm{10^{31}\ erg \ s^{-1}}$]  \\ 
\hline
Hard 		&  0.100		&    417		&	0.2	& 6.2			    	& 1.6 				& 1.6   \\
Hard 		&  0.235		&    417		&	2.4	& 33.9  				& 8.8 				& 9.0   \\ 
Moderate 	&  0.100		&    219		&	1.0	& 24.5  				& 2.4 				& 2.5    \\
Moderate 	&  0.235		&    219		&	1.0	& 68.4  		    	& 6.8 				& 6.9    \\ 
Soft    	&  0.173		&    151		&	0.2	& 50.4 					& 2.9 				& 2.9    \\
Soft 		&  0.235		&    151		&	0.7	& 59.3 					& 3.4 				& 3.4    \\ 
\hline\hline
\end{tabular}
\end{center}
\caption{Analysis results for the pulsar tail of \morla, taken at RA/Dec: (59.5313, 32.0372). The limits on luminosity assume a distance of 500~pc.}
\label{tab:veritas_upperlimits_morla} 
\end{table}

\begin{table}[htbp]
\begin{center}
\begin{tabular}{lcccccc}
\hline\hline
Spectral cuts	& $\theta$		& $\mathrm{E_{threshold}}$	& Significance & $95\%$ CL flux UL				& $95\%$ CL flux UL	& $95\%$ CL luminosity UL  \\
			&			&						& (Pre-Trials) &($>\mathrm{E_{threshold}}$)			&(1--10~TeV)			&(1--10~TeV) \\
			& [\degr]		&  [GeV]				&  [Sigma]	& [$\mathrm{10^{-13}\ cm^{-2}\ s^{-1}}$] 	& [$\mathrm{10^{-13}\ cm^{-2}\ s^{-1}}$]  & [$\mathrm{10^{31}\ erg \ s^{-1}}$]  \\ 
\hline
Hard			&  0.100		&    347			& 1.1		& 6.3  						& 1.2 				& 5.5   \\
Hard			&  0.235		&    347		    & 1.0		& 13.0  					& 2.6 				& 11.3   \\ 
Moderate		&  0.100		&    263			& 0.4		& 5.0  						& 0.65 				& 2.9   \\
Moderate		&  0.235		&    263			& 1.2		& 7.3 						& 0.95 				& 4.2   \\ 
Soft			&  0.173		&    166			& 3.5		& 3.8  						& 0.25 				& 1.1   \\
Soft			&  0.235		&    166			& 3.5		& 4.6  						& 0.30 				& 1.3   \\ 
\hline\hline
\end{tabular}
\end{center}
\caption{Analysis results for the pulsar tail of \mushroom, taken at RA/Dec: (59.6864, 54.1755). The limits on luminosity assume a distance of 1.04~kpc.}
\label{tab:veritas_upperlimits_mushroom} 
\end{table}


\begin{table}[htbp]
\begin{center}
\begin{tabular}{lcccccc}
\hline\hline
Spectral cuts	& $\theta$		& $\mathrm{E_{threshold}}$	& Significance & $95\%$ CL flux UL				& $95\%$ CL flux UL	& $95\%$ CL luminosity UL  \\
		&			&					& (Pre-Trials) &($>\mathrm{E_{threshold}}$)			&(1--10~TeV)		&(1--10~TeV) \\
		& [\degr]		&  [GeV]			& [Sigma]	& [$\mathrm{10^{-13}\ cm^{-2}\ s^{-1}}$] 	& [$\mathrm{10^{-13}\ cm^{-2}\ s^{-1}}$]  & [$\mathrm{10^{31}\ erg \ s^{-1}}$]  \\ 
\hline
Hard		&  0.100		&    501			& 0.5	& 4.1 						& 1.4 				& 10.6    \\
Hard		&  0.235		&    501			& 1.2	& 9.0  						& 3.1 				& 23.3    \\ 
Moderate	&  0.100		&    240			& 0.0	& 20.1  					& 2.4 				& 17.9    \\
Moderate	&  0.235		&    263			& 1.6	& 95.5 					    & 12.5 				& 93.7    \\ 
Soft		&  0.173		&    166			& 0.5	& 104.9  					& 6.9 				& 51.2    \\
Soft		&  0.235		&    182			& 1.1	& 195.3  					& 14.7 			    & 110.3   \\ 
\hline\hline
\end{tabular}
\end{center}
\caption{Analysis results for the pulsar tail of \psrj, taken at RA/Dec: (265.081, 9.96445). The limits on luminosity assume a distance of 1.36~kpc.}
\label{tab:veritas_upperlimits_j1740} 
\end{table}

\subsection{X-ray emission from the J0357+3205 tail}

Using {\sl XMM-Newton} data, \citet{Marelli2013} found that the spectrum of the tail fits a PL model with $\Gamma=2.07\pm0.08$ and $N_{\rm H}=(2.61\pm0.23)\times10^{21}$ cm$^{-2}$.
Fitting the {\sl CXO} data, we obtain compatible results, $\Gamma=2.11\pm0.12$,
$\mathcal{N}=(1.42\pm{0.17})\times10^{-4}$ photon s$^{-1}$ cm$^{-2}$ keV$^{-1}$ (at 1 keV)
and $N_{\rm H}=(3.5\pm0.5)\times10^{21}$ cm$^{-2}$ (with a reduced $\chi_\nu^2 = 1.01$, for $\nu=80$ d.o.f.). 
The entire tail was found to have a 0.5-8 keV luminosity $L=9\times10^{30}$ erg s$^{-1}$.
 \cite{DeLuca2011} found there is no statistically significant evidence of spectral evolution along the tail.  
It is worth noting that \citet{Klingler2016a,Klingler2016b} have shown  that some pulsar tails can exhibit a lack of spectral softening over parsec-scale distances from the pulsar.

\subsection{X-ray emission from the PSR B0355+54 tail and PSR J0359+5414 PWN}

\cite{Klingler2016b} reported the detailed analysis of {\sl CXO} observations of   PSR B0355+54 and its X-ray tail. In particular, the spectrum of the entire tail is well fitted ($\chi^2_\nu = 1.13$ for $\nu=135$ d.o.f.) by a single absorbed PL model with $\Gamma=1.74\pm0.08$, $\mathcal{N}=5.40\pm0.36$ photon cm$^{-2}$ s$^{-1}$ keV$^{-1}$ (at 1 keV), and  $ N_{\rm H} = 6.1\times10^{21}$ cm$^{-2}$.
No significant evidence of spectral cooling was seen between the near and far halves of the tail, as their photon indices were measured to be $\Gamma_{\rm near} = 1.72\pm0.10$ and $\Gamma_{\rm far} = 1.77\pm0.11$.
The entire tail was found to have a 0.5-8 keV luminosity $L=3.8\times10^{31}$ erg s$^{-1}$.


\cite{Zyuzin2018} reported the discovery of an X-ray PWN around $\gamma$-ray PSR J0359+5414 which is located just 5.5$'$ east of PSR B0355+54. 
The X-ray nebula of PSR J0359+5414 is much fainter than that of PSR B0355+54.  
The exposure-corrected images, shown in Figure \ref{fig-J0359}, reveal a faint and rather amorphous nebula that lacks any clear indication of substantial (supersonic) pulsar motion.
The total observed fluxes for the compact (within the polygon minus circle in Fig.~2) and extended (within the ellipse minus polygon in Fig.~2) fluxes are $8.6^{+0.6}_{-1.4}$  and $5.0^{+1.1}_{-1.8}\times10^{-15}$ erg s$^{-1}$ cm$^{-2}$ in 0.5-8 keV, respectively. 
 The corresponding best-fit PL photon indices are $\Gamma=1.7^{+0.4}_{-0.3}$ and $1.4^{+1.3}_{-0.9}$. 
The pulsar and compact nebula (CN) spectral fits suggest a hydrogen column density $N_{\rm H}\sim(2$-$6)\times10^{21}$ cm$^{-2}$.  The spectral fits of the much brighter B0355 PWN (located 5.5$'$ west of J0359, at parallax distance $d=1.04$ kpc) from the same set of \chandra\ observations yields $N_{\rm H,B0355}=(6.1\pm0.9)\times10^{21}$ cm$^{-2}$ \citep{Klingler2016b}. 
The similar $N_{\rm H}$ values suggest that the  distances to these two pulsars may be comparable. The combined (compact+extended) unabsorbed luminosity  of J0359's PWN is $\approx2.2\times10^{30}$ erg s$^{-1}$ in 0.5-8 keV at $d=1$ kpc.

\subsection{X-ray emission from the PSR J1740+1000 tail}
The combined MOS1/2 image (shown in Figure 1) reveals an X-ray tail extending southwest up to approximately 6$'$ from the pulsar position.
The initial segment of the tail appears slightly conical, with an opening angle of $\approx15^\circ$, but after about 3$'$ it appears to taper off and maintain a seemingly cylindrical shape.
The tail also appears to bend or curve slightly to the west with distance from the pulsar. 
Similar evidence of slight bending is also seen in the tails of PSRs B0355+54 (\citealt{Klingler2016b}; see also Figure 1) and J1741--2054 \citep{Auchettl2015}.

Based on the analysis of the PSR J1740+1000 spectrum, \citet{Kargaltsev2008} found the best-fit absorbing hydrogen column density $N_{\rm H}\approx1\times10^{21}$ cm$^{-2}$. 
However this value depends on the adopted spectral model, and the pulsar's spectrum requires multiple components for satisfactory fits.  
On the other hand, both the pulsar's dispersion measure and the HI column from radio data imply that the Galactic $N_{\rm H}$ in the direction of J1740 is $8\times10^{20}$ cm$^{-2}$ (see \citealt{He2013} and the HEAsoft/FTOOLs utility {\tt nH}), which is 
 close to that obtained by \citet{Kargaltsev2008}.
Therefore, while fitting the extended emission spectrum,  we fix\footnote{When left as a free parameter, the best-fit $N_{\rm H}$ goes to zero, which is unrealistic considering the pulsar's DM distance $d\sim1.4$ kpc.  In this fit the other parameters are as follows: $\mathcal{N}=(2.69\pm0.18)\times10^{-5}$ photons cm$^{-2}$ s$^{-1}$ keV$^{-1}$ (at 1 keV) and $\Gamma=1.54\pm0.06$, with a reduced $\chi^2_\nu = 0.97$ (for $\nu=201$ d.o.f.)}  $N_{\rm H}$ at  $10^{21}$ cm$^{-2}$.
The spectrum of the tail (extracted from the region shown in the left panel of Figure \ref{fig-pulsar-tails}) fits an absorbed PL with $\Gamma=1.75\pm0.04$ and a normalization $\mathcal{N}=(3.41\pm{0.09})\times10^{-5}$ photon s$^{-1}$ cm$^{-2}$ keV$^{-1}$ (at 1 keV), with a reduced $\chi^2_\nu = 1.04$ (for $\nu=202$ d.o.f.).  
The (absorbed) 0.3--10 keV flux $F_X=(1.93\pm0.06)\times10^{-13}$ erg cm$^{-2}$ s$^{-1}$ corresponds to an (unabsorbed) X-ray luminosity $L_X = (5.20\pm0.16)\times10^{31}$ erg s$^{-1}$ (at $d=1.4$ kpc), and an X-ray efficiency $\eta_X = 2.3\times10^{-4}$.

\section{DISCUSSION}

\begin{figure}
\epsscale{1.2}
\plotone{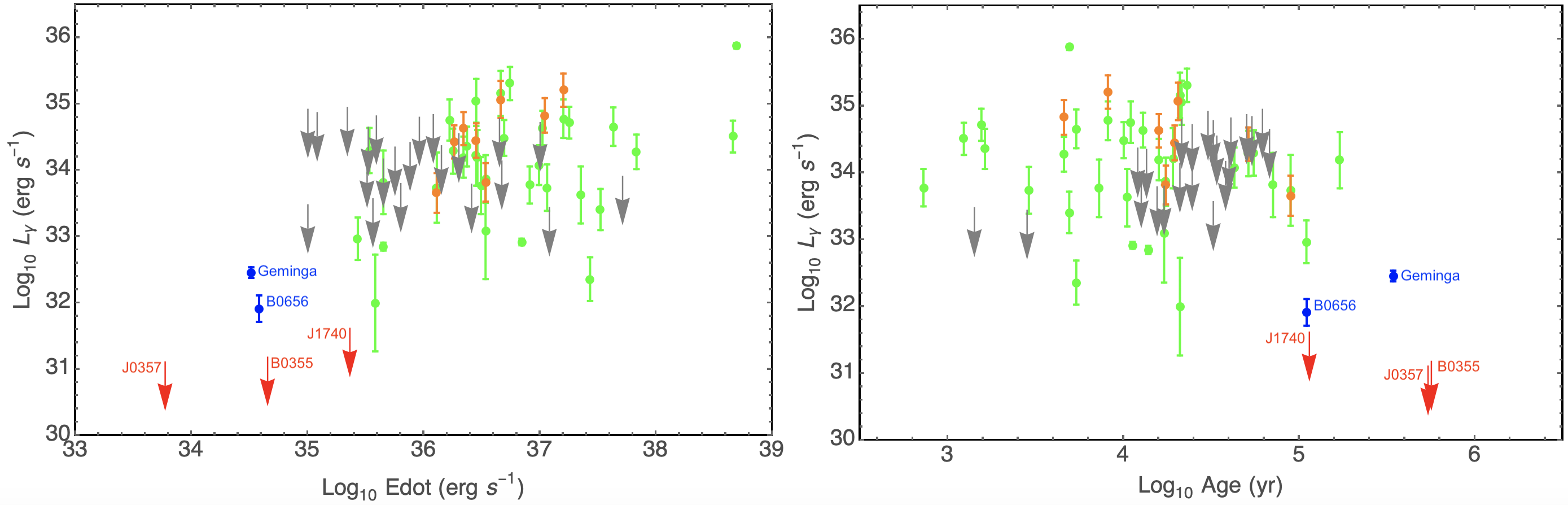}
\caption{Plots of 1--10 TeV luminosities $L_\gamma$ vs $\dot{E}$ and pulsar age.  The green points are confirmed TeV PWNe (from \citealt{Kargaltsev2010} and the HESS Galactic Plane Survey (HGPS), \citealt{HESS2017}, with values from the latter taking precedence for PWNe listed in both papers).  The orange points represent candidate PWNe from the HGPS with no incompatible measurements (see HGPS Table 4).  The gray arrows are the luminosity upper limits (also calculated at a 95\% CL) of pulsars whose PWNe were not detected in the HGPS.  The blue points represent the 1--10 TeV luminosities of Geminga and PSR B0656+14 (calculated from HAWC values for the 8-40 TeV range listed in \cite{2017Sci...358..911A} by assuming that the measured spectral slope remains the same in the 1-10 TeV range) and the red arrows are the luminosity upper limits of the targets observed in this study (using hard cuts).}
\label{fig-TeV-PWNe}
\end{figure}

Considering the entire population of TeV PWNe (see Figure \ref{fig-TeV-PWNe}), we see that the three tails discussed  here belong to pulsars that are older and have lower $\dot{E}$ compared to most of the others. 
The two exceptions are the nearby Geminga and B0656+14 pulsars. 
TeV sources associated with these PWNe have been detected by HAWC (\citealt{2017Sci...358..911A}, \citealt{Abeysekara2017}, \citealt{Baughman2015}). 

Due to their proximity as well as their large physical extent resolved by HAWC, the Geminga and B0656+14 PWNe take up a significant fraction of the VERITAS field of view. 
The standard source detection techniques common to IACTs (such as VERITAS), however, map gamma-ray sources using a round aperture of small angular size. 
This results in a limited sensitivity for two reasons: low surface brightness of the gamma-ray source relative to the cosmic ray background and difficulty obtaining a reliable background estimate from any location in the field of view. 
Thus, the current non-detection (see \citealt{2015arXiv150904224F}) of these sources by VERITAS does not objectively characterize its sensitivity to a distant TeV PWN of comparable physical size and luminosity.  

\begin{figure}
\epsscale{1.0}
\plotone{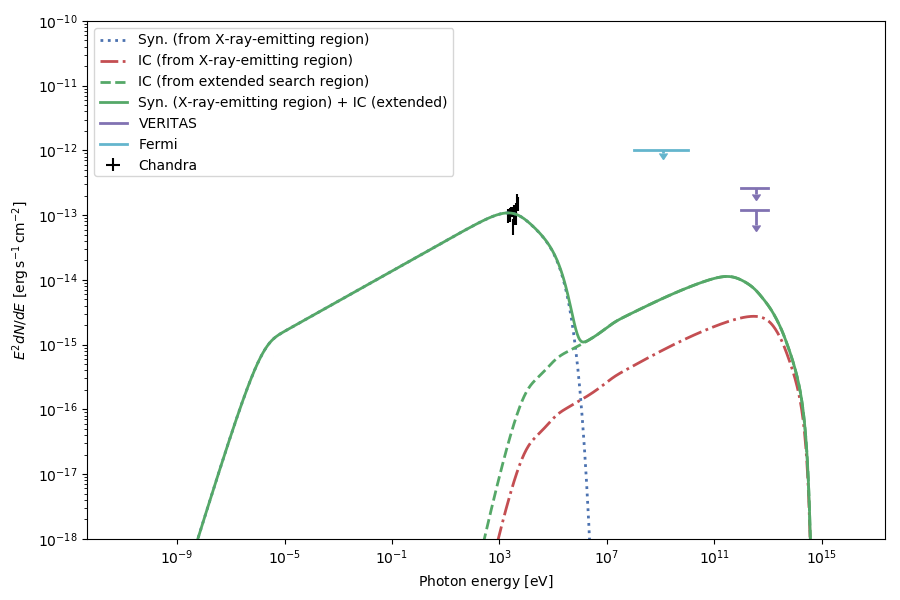}
\caption{The multiwavelength spectrum of the B0355 tail 
calculated from the pulsar tail model (see Appendix).
The blue (dotted) and red (dotted-dashed) lines represent the synchrotron and IC components (respectively) of the X-ray-emitting region of the tail (i.e., the regions shown in Figure \ref{fig-pulsar-tails}).  
The green (broad dashed) line represents the IC component of the extended tail 
 (see  text). 
Shown in solid green is the sum of the synchrotron component from the X-ray-emitting region of the tail and the IC component of the extended tail (which includes the X-ray-emitting region).
Also plotted are
 the measured \chandra\ spectrum of the X-ray tail (black data points), the {\sl Fermi}-LAT upper limit on the B0355 tail (cyan arrow), and the \veritas\ upper limits (purple arrows; the `hard cut' limits for the extraction region sizes of $r=0.1^{\circ}$ and $r=0.235^{\circ}$ are shown, with the more stringent upper limit  corresponding to the smaller region).
}
\label{fig-B0355-spec}
\end{figure}

\begin{figure}
\epsscale{1.0}
\plotone{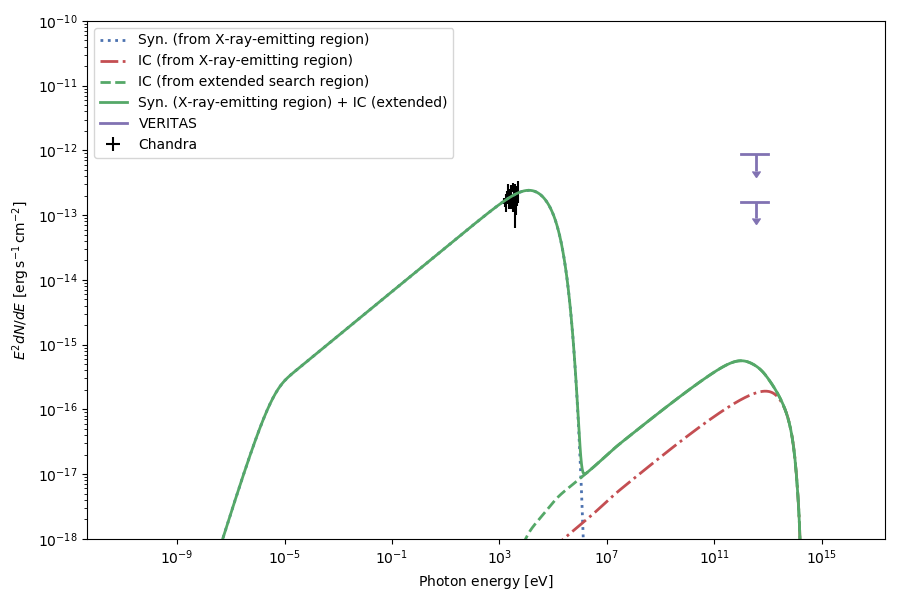}
\caption{The multiwavelength spectrum of the J0357 tail obtained from our pulsar tail model, with the same labeling scheme as the previous plot. (See text for discussion on the HAWC limit.)}
\label{fig-J0357-spec}
\end{figure}

\begin{figure}
\epsscale{1.0}
\plotone{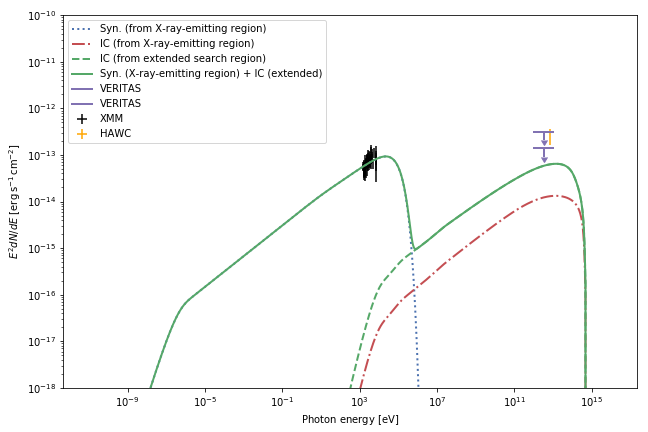}
\caption{The multiwavelength spectrum of the J1740 tail obtained from our pulsar tail model, with the same labeling scheme as the previous plot.}
\label{fig-J1740-spec}
\end{figure}

The VERITAS upper limit for the B0355 tail is already below the luminosity of the Geminga PWN inferred\footnote{Note that re-calculating the luminosity from the 8-40 TeV HAWC energy range to the 1-10 TeV range involved the assumption of a  constant spectral slope fixed at the value reported in \cite{2017Sci...358..911A}. The actual luminosity value can be smaller if the spectrum hardens toward lower energies.} from the HAWC measurements (see Figure 3).    
A more sophisticated analysis (see \cite{2019ICRC...36..616A}; beyond the scope of this paper) might further boost the  sensitivity to the extended PWN of B0355, as the round aperture used by VERITAS is not optimized for detection of an elongated tail.  The  non-detection of B0355's PWN may also be attributed to the differences between the Geminga and B0355 PWNe.
One such difference may be the morphology of  B0355's  PWN -- the long tail with fairly soft X-ray spectrum. 
Such a tail has not been detected for Geminga so far although the morphology of the compact X-ray nebula was interpreted as an SPWN \citep{Posselt2017}. 
{\sl eROSITA} should be able to tell whether Geminga lacks an extended tail or it is simply not detected due to its large size expected due to  Geminga's proximity to Earth.
Regardless, the Cherenkov Telescope Array (CTA), the next-generation IACT observatory, will have greatly improved sensitivity in the VHE range and should be able to easily detect pulsar tails like B0355's, thus enabling tests of the physical parameters associated with wind in pulsar tails.
To demonstrate this and to independently estimate probable TeV fluxes for the PWNe of the three pulsars considered in this paper, we performed modeling of the multiwavelength emission from these tails.  
 

The upper limits on the TeV flux from B0355 depend on assumptions about the size of the source. 
Determining the size of the tail is complicated. 
Firstly, the size of the TeV-emitting pulsar wind is expected to be larger than that of X-ray-emitting wind because of a factor of 10 longer cooling time for the TeV-emitting electrons.  
However, this is not the only factor that affects the length of the tail. 
The other two important factors are flow speed as a function of distance from the pulsar (which may be affected by the entrainment of the neutral atoms from the ambient medium; \citealt{Morlino2015}) and diffusion timescale for TeV-emitting electrons. 
Once the flow slows down significantly, approaching the pulsar speed (in the reference frame of the pulsar), diffusion is expected to start playing a dominant role, therefore turning the end of the tail into more of a spherical structure (unless the magnetic field preserves an ordered geometry on large scales and/or the external field is ordered). 
This could happen over the distance where the time to traverse the tail in the transverse direction, $r(z)/v(z)$, at the local bulk flow velocity, $v(z)$, becomes of the order of the diffusion timescale, $r(z)^2/D$ (where $D\simeq10^{26}(E_\gamma/1~{\rm TeV})^{1/2}(B/10~\mu{\rm G})^{-1}$ cm$^{2}$ s is the Bohm diffusion  coefficient of electrons that up-scatter the CMB photons to the gamma-ray energy $E_{\gamma}$). 
However, calculating the tail's disruption distance requires a better knowledge of $r(z)$, $v(z)$, and $B(z)$ than we currently have. Theoretical models suggest that the flow and magnetic field may be more ordered closer to the pulsar becoming more chaotic and turbulent further downstream (e.g., \citealt{2019MNRAS.484.4760B,2019ApJ...872...10X}). Observationally, these dependencies will become better constrained in the future if these tails are detected and resolved in TeV and radio.   Moreover, the TeV emission may extend even beyond the disruption distance, in which case 
the morphology would appear 
more isotropic than a collimated tail (which may be happening in the case of Geminga).  
Therefore, in this paper we adopt a heuristic approach to estimate possible sizes of TeV sources. 
We note that synchrotron cooling time in radio is much longer than that in X-rays, and that radio tails (when detected) extend up to 10$-$20 pc for highly supersonic pulsars like J1509--5850 \citep{Ng2010,Klingler2016a} and J1747--2958 \citep{Yusef-Zadeh2005, Klingler2018} which would correspond to $40'-80'$ at $d=1$ kpc (a typical distance to the three pulsars analyzed in this paper). 
Since at least B0355 appears to have a lower Mach number \citep{Klingler2016b}, we picked a compromise size of $20'-36'$ for the TeV emission search region.

To model the electron spectral energy distribution (SED) of the B0355 pulsar tail, we extend the one-dimensional (1D) model developed by \citet{Chen2006} to account for 
variations in tail properties (cross-section, magnetic field strength, and bulk flow velocity with distance from the pulsar), adiabatic energy losses, and the variable (decreasing) particle injection rate of the pulsar with time. 
The derivation of the model is explained in the Appendix. 
Using the electron SED parameters obtained from the fits to X-ray synchrotron spectrum from \chandra\ data (i.e., the SED responsible for the X-ray emission seen up to $\approx7'$ from the pulsar)
and the tail properties constrained from those fits (i.e., the bulk flow velocity and magnetic field strength estimates, see \citealt{Klingler2016b}),
we calculate the SED of the TeV-emitting particles up to 0.47$^\circ$ from the pulsar,
 and refer to these regions as the ``extended search regions'' (e.g., in Figure 4)  as opposed to  smaller regions corresponding to the  X-ray-emitting parts the tails (see Figure 1).
This length corresponds to the angular size of the region used for  extended TeV emission search (see Table \ref{tab:veritas_upperlimits_mushroom} and Section 2.1) and 
 is also comparable to the recently-detected extended TeV emission from the Geminga pulsar (see \citealt{Abeysekara2017} \& \citealt{Linden2017}).
We used the Python package {\tt Naima} \citep{Zabalza2015} to calculate the synchrotron and IC 
emission from the volume-averaged electron SED (see Appendix), and present the multiwavelength spectrum of B0355 in Figure \ref{fig-B0355-spec}. 
The IC emission is calculated for the combined contributions of the CMB, thermal dust emission (30 K, $4.8\times10^{-13}$ erg cm$^{-3}$), and starlight (4000 K, $8.0\times10^{-13}$ erg cm$^{-3}$).
We see that an observation more sensitive by a factor of 10 may be able to detect TeV emission. 
The modeling also demonstrates that, if the TeV spectrum is measured in addition to X-ray properties of the tail, one can put more stringent constraints on dependencies of the magnetic field strength and bulk flow velocity on the distance from the pulsar (see Appendix). 

Unfortunately, despite the detailed X-ray information available for B0355,  modeling its TeV emission (if it is detected) is  complicated by the presence of another middle-aged pulsar, J0359, just $5.5'$ away. Having an angular resolution of $\sim1'$ in the VHE range will be critical to establish the origin of any TeV emission. Indeed, the comparable absorbing hydrogen column density, $N_{\rm H}\sim(0.2 - 0.6)\times10^{22}$ cm$^{-1}$ for J0359 compared to $N_{\rm H}=0.6\times10^{22}$ cm$^{-2}$ for B0355,  suggests that J0359 is located at a similar distance, if not closer than B0355. Given the relatively high $\dot{E}=1.3\times10^{36}$ erg s$^{-1}$  (compared to $4.5\times10^{34}$ erg s$^{-1}$) the X-ray efficiency of J0359 PWN is very low, $\eta_X=L_X/\dot{E}\sim10^{-6}$ for $d=1$ kpc (compared to $\eta_X=8\times10^{-4}$ for B0355), making it one of the least efficient in X-rays. It is currently unclear why X-ray  efficiencies of PWNe vary so much \citep{KP08} and whether there is a correlation between X-ray and TeV efficiencies for PWNe (see Fig.~4 in \citealt{Kargaltsev2013}). The non-detection in TeV
 suggests that   the  PWN of J0359 is also very inefficient in terms of its TeV emission.


We also repeated the same steps  as above (see Appendix for details) to  model  the multiwavelength spectra of  J0357 and J1740 tails (assuming a constant tail radius, flow speed, and magnetic field  because we do not have as much information as for the B0355 tail from X-ray data), and obtain both the particle SED in the $8\farcm5$ X-ray-emitting part of the tail producing the observed synchrotron emission (11$z_0$; see appendix for definition of $z_0$), and the SED of the TeV-emitting particles up to a tail length of 30$'$ (50$z_0$, comparable to the angular size of the region of sky analyzed, see Tables \ref{tab:veritas_upperlimits_morla} and \ref{tab:veritas_upperlimits_j1740} along with Figures \ref{fig-J0357-spec} and \ref{fig-J1740-spec}).  The modeling suggests that detecting J0357 in TeV would be much more difficult compared to the other two pulsars.

TeV emission from an unresolved source located near the J1740 tail (at the HAWC resolution it can be considered as a match within the $2\sigma$ positional uncertainty of the 2HWC source) was recently  reported in the 3HWC catalog \citep{Albert2020} as 3HWC J1739+099. However, the source significance is fairly low  (just above the threshold chosen for the 3HWC catalog).
The source's differential flux at 7 TeV (plotted in Figure \ref{fig-J1740-spec}), $2.5\times10^{-13}$ erg s$^{-1}$ cm$^{-2}$ is about a factor of 3-4 higher than predicted by our model.
This can be attributed to the fact that we do not reliably know the shape and size of the tail (which also affects other assumptions and variables used in the model, like the bulk flow speed magnetic field's dependence on distance from the pulsar).
There may also be a component from electrons that are still radiating in TeV in the tail's vicinity, but that are not necessarily in the tail anymore (in which case they are not subject to adiabatic energy losses and synchrotron losses due to the tail's magnetic field).
At some point the tail flow is disrupted or ceases, and electrons diffuse out into the ISM, leading to so-called TeV halos, such as the one around Geminga, for example (this component is not modeled here).


\section{Summary and Conclusions}


We analyzed VERITAS observations and reported the upper limits on the TeV fluxes of three supersonically moving pulsars with long tails seen in X-rays. TeV emission from SPWNe continues to remain elusive, although our modeling suggests that any TeV emission could be detected with only a factor of 2-3 improved flux sensitivity, for J1740's tail, and a factor of 10, for B0355's tail, in deeper observations. In fact, when this paper was nearing completion the HAWC team reported a detection, albeit at a modest significance, of a VHE source located near the most promising tail, associated with PSR J1740+1000.  We expect that CTA will be able to show conclusively whether the TeV emission from SPWNe is substantially different from that of slower moving or younger pulsars still residing within their host SNRs. 


\acknowledgements

This research is supported by grants from the U.S. Department of Energy Office of Science, the U.S. National Science Foundation and the Smithsonian Institution, by NSERC in Canada, and by the Helmholtz Association in Germany. We acknowledge the excellent work of the technical support staff at the Fred Lawrence Whipple Observatory and at the collaborating institutions in the construction and operation of the instrument. The work of OK and IV  was supported by NASA Astrophysics Data Analysis Program award 80NSSC19K0576 and NASA {\sl XMM-Newton} award 80NSSC18K0636.  OK work on this project  was also supported by the  Chandra Award TM8-19005B issued by the Chandra X-ray Center, which is operated by the Smithsonian Astrophysical Observatory for and on behalf of NASA under contract NAS8-03060.


This research used resources of the National Energy Research Scientific Computing Center, a DOE Office of Science User Facility supported by the Office of Science of the U.S. Department of Energy under Contract No. DE-AC02-05CH11231.

\software{Naima \citep{Zabalza2015}, CIAO (v4.11; \citealt{Fruscione2006}), XSPEC (v12.10.1; \citealt{Arnaud1996}), SAS (v16.1; \citealt{Gabriel2004}), VEGAS \citep{Daniel2008}, Eventdisplay \citep{2017ICRC...35..747M} } 


\appendix
We construct a pulsar tail model in order to calculate the electron SED and the corresponding IC and synchrotron spectra as a function of the distance from the pulsar.
We compare these model predictions with the obtained TeV upper limits.

The simple one-dimensional (1D) model developed by \citet{Chen2006} describes the impact of synchrotron losses on the electron SED as electrons (injected with some initial power-law SED at the TS) move along a cylindrical tail with constant bulk velocity. 
In that model, the tail's radius and magnetic field are assumed to be constant. 
We relax the model assumptions by allowing for a variable (with the distance from pulsar) magnetic field $B$, particle bulk flow velocity $v$, and tail radius $r$, 
\begin{equation}
B=B(z)=B_0\left(\frac{z}{z_0} \right)^\beta		\qquad	v=v(z)=v_0 \left(\frac{z}{z_0} \right)^\alpha		\qquad	r=r(z)=r_0 \left(\frac{z}{z_0} \right)^\gamma
\end{equation}
where $z_0$ is the distance from the pulsar at which particles are injected into the tail, $z_0$,  $z>z_0$ is the distance along the tail, and $B_0$,  $v_0$, and $r_0$ are  the initial values of the magnetic field, velocity, and radius, respectively. The constants  $\beta$, $\alpha$, and $\gamma$ describe how these quantities change with distance, respectively.
We pick the injection distance $z_0$ such that it corresponds with the ``beginning'' of the X-ray tail (i.e., the boundary between the compact nebula head and the beginning of the tail).
The energy loss rate of an electron with energy $\mathcal{E}$ (due to both synchrotron and adiabatic losses) is given by
\begin{equation}
\frac{d\mathcal{E}}{dt} = -c_1 B^2 \mathcal{E}^2 - \frac{1}{3}\left( \frac{\alpha+2\gamma}{z_0} \right) v_0 \left( \frac{z}{z_0} \right)^{\alpha-1}\mathcal{E}
\end{equation}
where $c_1=4e^4/9m_e^4 c^7$.
Solving the differential equation yields
\begin{equation}
\mathcal{E}(z,\mathcal{E}_0,z_0) = \frac{\mathcal{E}_0 \left( \frac{z}{z_0} \right)^{-\lambda_1}}{1 + \frac{\mathcal{E}_0}{\epsilon \lambda_2} \left[ \left( \frac{z}{z_0} \right)^{-\lambda_2} -1 \right] }
\end{equation}
where $\epsilon = v_0 / (c_1 B_0^2 z_0)$, $\lambda_1 = (\alpha + 2\gamma)/3$, $\lambda_2 = 2\beta - \alpha - \lambda_1 + 1$,
and $\mathcal{E}_0$ is the initial energy of the particle.  
Thus, the  number density of particles with energy $\mathcal{E}$ at distance $z$, $n_e(\mathcal{E},z)$, can be found from the continuity condition, $n_e(\mathcal{E},z)\ \pi r^2 v = {\rm constant}$.
Assuming the injected particle distribution to be $n_e(\mathcal{E}_0, z=z_0) = K \mathcal{E}_0^{-p}$ $(\mathcal{E}_{\rm 0,min} \leq \mathcal{E}_0 \leq \mathcal{E}_{\rm 0,max})$, one obtains
\begin{equation}
n_e(\mathcal{E},z) = K \mathcal{E}^{-p} \frac{ \left( \frac{z}{z_0} \right)^{-2\gamma - \alpha - \lambda_1} }{ \left\{ \left( \frac{z}{z_0} \right)^{-\lambda_1} - \frac{E}{\epsilon \lambda_2} \left[ \left( \frac{z}{z_0} \right)^{-\lambda_2} -1 \right]  \right\}^{2-p} }  \theta(\mathcal{E}_{\rm max} - \mathcal{E} ),
\end{equation}
where, $\theta(\mathcal{E}_{\rm max} - \mathcal{E} )$ is the Heaviside step function,  $\mathcal{E}_\text{max}=\mathcal{E}(z,\mathcal{E}_\text{0,max},z_0)$, $\mathcal{E}_{\rm 0,min}=0$, and $\mathcal{E}_{\rm 0,max}$ is assumed to be the energy associated with the maximum potential drop between the pole and the light cylinder,  $\Delta\Phi=(3\dot{E}/2c)^{1/2}$  \citep{GoldreichJulian1969}.

The particle injection rate is a function of $\dot{E}$, which decreases as the pulsar ages.  Thus, at increasing distances $z$, the associated injection rate at the corresponding time of injection (in the past) $t$ was higher than the current injection rate:  
\begin{equation}
\dot{E}(t) = \frac{\dot{E}_0}{\left(1+ \frac{\tau-t}{\tau_0} \right)^{\frac{n+1}{n-1}}}
\end{equation}
where $\tau$ is the current age of the pulsar, $t$ is the time since injection, $\tau_0$ is the spin-down timescale, and $n$ is the braking index (data from the HESS Galactic Plane Survey suggests that $\tau_0$ ranges between $10^{2.5}$ and $10^{3.5}$ yr; see \citealt{HESS2017}; canonically, a simple spinning dipole corresponds to breaking index $n=3$).
One can take the variable injection rate into account in Equation (4) by multiplying it with $1/(1+(\tau-t)/\tau_0)$ (assuming $n= 3$ and particle injection rate $\propto \dot{E}^{1/2}$; see \cite{2012SSRv..173..341A}), after rewriting $t=dz/v(z)$ and integrating to $z$.
Thus, the volume-averaged electron SED can be obtained by integrating $n_e(\mathcal{E},z)$ over the volume of the tail.
For the tail lengths, we picked $z_{\rm max}$ values that correspond to the size of the extended TeV search region ($0.47$\degr, see above).

To model the B0355 tail, we use the following parameters: $B_0=20\ {\rm \mu G}$, $z_0=1\times10^{18}$ cm, $v_0=2,400$ km s$^{-1}$, $r_0=4.6\times10^{17}$ cm, $\gamma=1$ (conical tail), $\beta=-0.35$, $\alpha=-0.65$ (constrained from $\beta=-\alpha-\gamma$), $p\equiv 2\Gamma-1=2.48$, $\mathcal{E}_{\rm 0,max}=580$ erg (set by the maximum accelerating potential of the pulsar's polar cap, $\Phi=(3\dot{E}/2c)^{1/2}$), and $\tau=560$ kyr.
The magnetic field parameters, flow speed parameters, and particle spectrum $p$ are consistent with those reported by \citet{Klingler2016b}.
The $z_0=1\times10^{18}$ cm corresponds to the distance at which the tail begins, downstream of the compact nebula.
For this tail (as well as the others), we model the tail out to a distance corresponding to $0.47^\circ$, as this angular distance corresponds to the angular size of the largest VERITAS search region  (see Section 2.1 and Tables 2--4).

To model the J0357 tail, we use $d=500$ pc (the distance inferred by \citealt{DeLuca2011}), 
$z_0=3.6\times10^{17}$ cm (the distance at which the tail becomes visible, $0\farcm8$), 
$z_{\rm max} = 50 z_0$,
$r_0=2.7\times10^{17}$ cm (corresponding to the tail's $0\farcm6$ radius), 
and $v=v_0=15,000$ km s$^{-1}$ and $B=B_0=50\ {\rm \mu G}$ (see \citealt{DeLuca2011} and \citealt{Marelli2013}).  
In this case we assume a cylindrical tail (constant radius, $\gamma=0$) with constant flow speed and magnetic field strength ($\alpha=\beta=0$), 
$p=2.64$ ($\Gamma=1.82$, obtained by fitting the entire tail spectrum with $N_{\rm H}=0.21\times10^{22}$ cm$^{-2}$: the maximum Galactic $N_{\rm H}$ in the direction of J0357, obtained by \citealt{Marelli2013}),
and set $\mathcal{E}_{\rm 0,max}=260$ erg and $\tau=540$ kyr.  
Since the quantities here are less constrained than those of B0355 (e.g., $B$ due to lack of multiwavelength observations, and $d$ being only inferred), this model can only serve as a crude approximation in this case.

To model the J1740 tail, we use the dispersion measure distance $d\sim1.4$ kpc, $z_0=8.4\times10^{17}$ cm (40$''$; we select this distance because it is the distance at which the bright pulsar's PSF (in the {\sl XMM} data) no longer contaminates the tail emission), $r_0=8.4\times10^{17}$ cm, and $\gamma=0$ (cylindrical tail; thus $\alpha=\beta=0$).
For the tail length we used $z_{\rm max} = 40 z_0$.
To estimate the magnetic field strength, we use Eqn.\ (2) from \citet{Klingler2016a}, which is a generalization of Eqn.\ (7.14) from \citet{Pacholczyk1970} for the case of arbitrary magnetization $\sigma$,
\begin{equation}
B = \left[ \frac{L(\nu_m,\nu_M) \sigma}{\mathcal{A}V} \frac{\Gamma-2}{\Gamma-1.5} \frac{\nu_1^{1.5-\Gamma} - \nu_2^{1.5\Gamma}}{\nu_m^{2-\Gamma} - \nu_M^{2-\Gamma}} \right]^{2/7}.
\end{equation}
Here, $\nu_1$ and $\nu_2$ are the characteristic synchrotron frequencies ($\nu_{\rm syn} \simeq eB\gamma^2 / 2\pi mc$ with $\gamma=\mathcal{E}/mc^2$) corresponding to the upper and lower energies of the electron SED $(dN_e / d\gamma \propto \gamma^{-p} \propto \gamma^{-2\Gamma+1}$; $\gamma_1 < \gamma < \gamma_2)$, and $\mathcal{A}=2^{1/2} e^{7/2} / 18 \pi^{1/2} m_e^{5/2} c^{9/2} $.
The radiating volume $V$ can be estimated by assuming a 6$'$-long cylindrical tail of radius $r_0$, in which case $V=1.66\times10^{55}$ cm$^3$.
The photon index $\Gamma$ and luminosity $L$ are measured over the range $\nu_m < \nu < \nu_M$ (0.3--10 keV; see Section 3.2).
The boundary frequencies $\nu_1$ and $\nu_2$ are unknown, so we choose arbitrary yet plausible values $h\nu_1=3$ eV and $h\nu_2 = 20$ keV.
We estimate $B=B_0=5.1\sigma^{2/7}$ $\mu$G, a reasonable value given the pulsar's offset from the Galactic plane.
Extending the range of the boundary frequencies to $h\nu_1=3$ eV and $h\nu_2=60$ keV and varying $\Gamma$ within its 1$\sigma$ uncertainty results in only minor changes, with the resulting estimates of $B$ ranging from 5.0--6.4 $\mu$G.
We also set $\mathcal{E}_{\rm 0,max}=1600$ erg and $\tau=114$ kyr, and assume $v=v_0=5,000$ km/s. The latter could be fitted  when better TeV data become available.  


\begin{thebibliography}{99}



\bibitem[Abeysekara et al.(2017a)]{2017Sci...358..911A} Abeysekara, A.~U., Albert, A., Alfaro, R., et al.\ 2017a, Science, 358, 911
\bibitem[Abeysekara et al.(2017b)]{Abeysekara2017} Abeysekara, A.~U., Albert, A., Alfaro, R., et al.\ 2017b, \apj, 843, 40 
\bibitem[Abeysekara(2019)]{2019ICRC...36..616A} Abeysekara, A.\ 2019, 36th International Cosmic Ray Conference (ICRC2019), 36, 616
\bibitem[Abdo \etal(2009)]{Abdo2009}Abdo, A.~A., \etal\ 2009, Science, 325, 5942, 840
\bibitem[Abramowski et al.(2012)]{Abramowski2012} Abramowski, A., Acero, F., Aharonian, F., et al.\ 2012, \aap, 548, A38. doi:10.1051/0004-6361/201219919
\bibitem[Acciari \etal(2008)]{Acciari2008}Acciari, V.~A., \etal\ 2008, ApJ, 679, 1427
\bibitem[Acero \etal(2015)]{Acero2015}Acero, F., \etal\ 2015, ApJSS, 218, 23
\bibitem[Aharonian \etal(1997)]{Aharonian1997}Aharonian, F.~A., \etal\ 1997, APh, 6, 343
\bibitem[Aharonian et al.(2006)]{Aharonian2006} Aharonian, F., Akhperjanian, A.~G., Bazer-Bachi, A.~R., et al.\ 2006, \aap, 448, L43. doi:10.1051/0004-6361:200600014
\bibitem[Albert et al.(2020)]{Albert2020} Albert, A., Alfaro, R., Alvarez, C., et al.\ 2020, arXiv:2007.08582
\bibitem[Arnaud(1996)]{Arnaud1996} Arnaud, K.~A.\ 1996, Astronomical Data Analysis Software and Systems V, 101, 17
\bibitem[Arons(2012)]{2012SSRv..173..341A} Arons, J.\ 2012, \ssr, 173, 341. doi:10.1007/s11214-012-9885-1
\bibitem[Auchettl et al.(2015)]{Auchettl2015} Auchettl, K., Slane, P., Romani, R.~W., et al.\ 2015, \apj, 802, 68 
\bibitem[Bandiera (2008)]{Bandiera2008}Bandiera, R. 2008, A\&A, 490, L3
\bibitem[Barkov et al.(2019)]{2019MNRAS.484.4760B} Barkov, M.~V., Lyutikov, M., \& Khangulyan, D.\ 2019, \mnras, 484, 4760. doi:10.1093/mnras/stz213
\bibitem[Baughman et al.(2015)]{Baughman2015} Baughman, B.~M., Wood, J., \& for the HAWC Collaboration 2015, arXiv:1508.03497 
\bibitem[Berge \etal(2007)]{Berge2007}Berge, D., Funk, S., \& Hinton, J. 2007, A\&A, 466, 1219
\bibitem[Bykov et al.(2017)]{Bykov2017} Bykov, A.~M., Amato, E., Petrov, A.~E., Krassilchtchikov, A.~M., \& Levenfish, K.~P.\ 2017, \ssr, 207, 235 
\bibitem[Cash(1979)]{Cash1979} Cash, W.\ 1979, \apj, 228, 939 
\bibitem[Chatterjee \etal(2004)]{Chatterjee2004}Chatterjee, S., \etal\ 2004, ApJ, 604, 339
\bibitem[Chen \etal(2006)]{Chen2006}Chen, Y., Wang, Q.~D., Gotthelf, E.~V., \etal\ 2006, ApJ, 651, 237 
\bibitem[Chevalier \& Reynolds(2011)]{Chevalier2011} Chevalier, R.~A. \& Reynolds, S.~P.\ 2011, \apjl, 740, L26. doi:10.1088/2041-8205/740/1/L26
\bibitem[Clark \etal(2017)]{Clark2017}Clark, C.~J., Pletsch, H.~J., Guillemot, L. \etal\ 2017, ApJ, 834, 106C
\bibitem[Cogan (2008)]{Cogan2008}Cogan, P. 2008, in Proc 30th ICRC, Merida, Mexico, 3, 1385
\bibitem[Daniel (2008)]{Daniel2008}Daniel, M.~K. 2008, in Proc 30th ICRC, Merida, Mexico, 3, 1325
\bibitem[Deller et al.(2019)]{Deller2019} Deller, A.~T., Goss, W.~M., Brisken, W.~F., et al.\ 2019, \apj, 875, 100 
\bibitem[De~Luca \etal(2011)]{DeLuca2011}De~Luca, A., \etal\ 2011, ApJ, 733, 104
\bibitem[De~Luca \etal(2013)]{DeLuca2013}De~Luca, A., \etal\ 2013, ApJL, 765, L19
\bibitem[Gabriel et al.(2004)]{Gabriel2004} Gabriel, C., Denby, M., Fyfe, D.~J., et al.\ 2004, Astronomical Data Analysis Software and Systems (ADASS) XIII, 314, 759
\bibitem[Goldreich \& Julian(1969)]{GoldreichJulian1969} Goldreich, P., \& Julian, W.~H.\ 1969, ApJ, 157, 869
\bibitem[Flinders(2015)]{2015arXiv150904224F} Flinders, A.\ 2015, arXiv:1509.04224
\bibitem[Fruscione et al.(2006)]{Fruscione2006} Fruscione, A., McDowell, J.~C., Allen, G.~E., et al.\ 2006, \procspie, 6270, 62701V. doi:10.1117/12.671760
\bibitem[H.~E.~S.~S.~Collaboration et al.(2017)]{HESS2017} H.~E.~S.~S.~Collaboration; Abdalla, H., \etal\ 2017, arXiv:1702.08280
\bibitem[Halpern \etal(2013)]{Halpern2013}Halpern, J.~P., \etal\ 2013, ApJ, 778, 2
\bibitem[He et al.(2013)]{He2013} He, C., Ng, C.-Y., \& Kaspi, V.~M.\ 2013, \apj, 768, 64 
\bibitem[Helfand (1983)]{Helfand1983}Helfand, D.~J. 1983, in Supernova Remnants and their X-ray Emission, 471
\bibitem[Hillas (1985)]{Hillas1985}Hillas, M. 1985, in Proc 19th ICRC, La Jolla, IL, 3, 445
\bibitem[Hobbs et al.(2005)]{Hobbs2005} Hobbs, G., Lorimer, D.~R., Lyne, A.~G., \& Kramer, M.\ 2005, \mnras, 360, 974 
\bibitem[Hofmann \etal(1999)]{Hofmann1999}Hofmann, W., \etal\ 1999, APh, 12, 135
\bibitem[Holder \etal(2006)]{Holder2006}Holder, J., \etal\ 2006, APh, 25 (6), 391
\bibitem[Hui et al.(2015)]{Hui2015} Hui, C.~M., Zhou, H., \& for the HAWC Collaboration 2015, arXiv:1508.07391 
\bibitem[Kargaltsev \etal(2008)]{Kargaltsev2008} Kargaltsev, O., \etal\ 2008, ApJ, 684, 1
\bibitem[Kargaltsev \& Pavlov(2008)]{KP08} Kargaltsev, O., \& Pavlov, G.~G.\ 2008, 40 Years of Pulsars: Millisecond Pulsars, Magnetars and More, 171
\bibitem[Kargaltsev \& Pavlov(2010)]{Kargaltsev2010} Kargaltsev, O., \& Pavlov, G.~G.\ 2010, X-ray Astronomy 2009; Present Status, Multi-Wavelength Approach and Future Perspectives, 1248, 25 
\bibitem[Kargaltsev \etal(2013)]{Kargaltsev2013} Kargaltsev, O., Rangelov, B., \& Pavlov, G.~G.\ 2013, arXiv e-prints, arXiv:1305.2552
\bibitem[Kargaltsev et al.(2017)]{Kargaltsev2017} Kargaltsev, O., Klingler, N., Chastain, S., \& Pavlov, G.~G 2017, arXiv:1711.02656 
\bibitem[Krawczynski \etal(2006)]{Krawczynski2006}Krawczynski, H., \etal\ 2006, APh, 25, 380
\bibitem[Klingler \etal(2016a)]{Klingler2016a} Klingler, N., Kargaltsev, O., Rangelov, B., et al.\ 2016a, ApJ, 828, 70 
\bibitem[Klingler \etal(2016b)]{Klingler2016b} Klingler, N., Rangelov, B., Kargaltsev, O., \etal\ 2016b, ApJ, 833, 253 
\bibitem[Klingler et al.(2018)]{Klingler2018} Klingler, N., Kargaltsev, O., Pavlov, G.~G., et al.\ 2018, \apj, 861, 5
\bibitem[Li \& Ma(1983)]{Li1983}Li, T.-P. \& Ma, Y.-Q. 1983, ApJ, 272, 317
\bibitem[Linden \etal(2017)]{Linden2017} Linden, T., Auchettl, K., Bramante, J., \etal\ 2017, arXiv:1703.09704 
\bibitem[Maier \& Holder(2017)]{2017ICRC...35..747M} Maier, G. \& Holder, J.\ 2017, 35th International Cosmic Ray Conference (ICRC2017), 301, 747
\bibitem[Manchester et al.(2005)]{Manchester2005} Manchester, R.~N., Hobbs, G.~B., Teoh, A., et al.\ 2005, \aj, 129, 1993
\bibitem[Marelli \etal(2013)]{Marelli2013}Marelli, M., \etal\ 2013, ApJ, 765, 36
\bibitem[Marelli \etal(2016)]{Marelli2016}Marelli, M., \etal\ 2016, ApJ, 819, 40
\bibitem[Morlino et al.(2015)]{Morlino2015} Morlino, G., Lyutikov, M., \& Vorster, M.\ 2015, \mnras, 454, 3886
\bibitem[McLaughlin \etal(2000)]{McLaughlin2000} McLaughlin, M., \etal\ 2000, in Pulsar Astronomy - 2000 and Beyond, 41
\bibitem[McLaughlin \etal(2002)]{McLaughlin2002} McLaughlin, M.~A., \etal\ 2002, ApJ, 564, 333
\bibitem[McGowan \etal(2006)]{McGowan2006}McGowan, K.~E., \etal\ 2006, ApJ, 647, 1300
\bibitem[Ng et al.(2010)]{Ng2010} Ng, C.-Y., Gaensler, B.~M., Chatterjee, S., et al.\ 2010, \apj, 712, 596
\bibitem[Ng et al.(2012)]{Ng2012} Ng, C.-Y., Bucciantini, N., Gaensler, B.~M., et al.\ 2012, \apj, 746, 105
\bibitem[Pacholczyk(1970)]{Pacholczyk1970} Pacholczyk, A.~G.\ 1970, Series of Books in Astronomy and Astrophysics, San Francisco: Freeman, 1970, 
\bibitem[Park \etal(2015)]{Park2015}Park, N. 2015, in Proc 34th ICRC, The Hague, Netherlands, ArXiv e-prints 1508.07070
\bibitem[Pavan \etal(2014)]{Pavan2014}Pavan, L. \etal\ 2014, Int. J. Mod. Phys. Conf. Ser., 28, 1460172
\bibitem[Pavan et al.(2014)]{Pavan2014} Pavan, L., Bordas, P., P{\"u}hlhofer, G., et al.\ 2014, \aap, 562, A122
\bibitem[Posselt et al.(2017)]{Posselt2017} Posselt, B., Pavlov, G.~G., Slane, P.~O., et al.\ 2017, \apj, 835, 66
\bibitem[Rolke \etal(2005)]{Rolke2005}Rolke, W. A., \etal\ 2005, Nucl.\ Instrum.\ Meth. A551, 493
\bibitem[Saz~Parkinson \etal(2010)]{SazParkinson2010}Saz~Parkinson, P. M., \etal\ 2010, ApJ, 725, 571
\bibitem[Seward \& Wang(1988)]{Seward1988}Seward, F.~D. \& Wang, Z.-R. 1988, ApJ, 332, 199
\bibitem[Slane (1994)]{Slane1994}Slane, P. 1994, ApJ, 437, 458
\bibitem[Slane et al.(2018)]{Slane2018} Slane, P., Lovchinsky, I., Kolb, C., et al.\ 2018, \apj, 865, 86. doi:10.3847/1538-4357/aada12
\bibitem[Tepedelenli{\v g}lu \& {\"O}gelman(2007)]{Tepedelenliglu2007}Tepedelenli{\v g}lu, E. \& {\"O}gelman, H. 2007, ApJ, 658, 1183
\bibitem[Verbiest \etal(2012)]{Verbiest2012}Verbiest, J. P. W., \etal\ 2012, ApJ, 755, 39
\bibitem[Verbunt et al.(2017)]{Verbunt2017} Verbunt, F., Igoshev, A., \& Cator, E.\ 2017, \aap, 608, A57
\bibitem[Wilms \etal(2000)]{Wilms2000} Wilms, J., Allen, A., \& McCray, R.\ 2000, \apj, 542, 914 
\bibitem[Xu et al.(2019)]{2019ApJ...872...10X} Xu, S., Klingler, N., Kargaltsev, O., et al.\ 2019, \apj, 872, 10
\bibitem[Yusef-Zadeh \& Gaensler(2005)]{Yusef-Zadeh2005} Yusef-Zadeh, F., \& Gaensler, B.~M.\ 2005, Advances in Space Research, 35, 1129
\bibitem[Zabalza(2015)]{Zabalza2015} Zabalza, V.\ 2015, arXiv:1509.03319 
\bibitem[Zyuzin et al.(2018)]{Zyuzin2018} Zyuzin, D.~A., Karpova, A.~V., \& Shibanov, Y.~A.\ 2018, \mnras, 476, 2177


\end{thebibliography}
\end{document}